\documentstyle[12pt,psfig]{article}

\begin{document}

\begin{titlepage}
\begin{center}

{\Large\bf{The High Density Effects in Heavy Quark Production at
$pA$ Colliders}}
\\[5.0ex]
{\Large\it{A. L. Ayala Filho
$^{1,*}$\footnotetext{$^{*}$E-mail:ayala@ufpel.tche.br}}}
 {\it and}
{ \Large \it{ V. P.  Gon\c{c}alves
$^{1,2,**}$\footnotetext{$^{**}$E-mail:barros@ufpel.tche.br} }}
\\[1.5ex]
 {\it  $^1$ Instituto de F\'{\i}sica e Matem\'atica, Univ. Federal
de Pelotas}\\ {\it Caixa Postal 354, 96010-090, Pelotas, RS,
BRAZIL}
\\[1.5ex]
 {\it  $^2$ Universidade Estadual do Rio Grande do Sul - UERGS,} \\
 {\it Borges de Medeiros, 1501,
  90119-900, Porto Alegre,  RS,
BRAZIL}
\\[5.0ex]
\end{center}

{\large \bf Abstract:}  In this paper we investigate the role of
the high density effects in the heavy quark production cross
section in $pA$ processes at RHIC and LHC. We use, as  initial
condition, a gluon distribution consistent with fixed target
nuclear data and the Glauber-Mueller approach to describe the high
density effects. We show that this process can be used as a probe
of the presence of the high density effects. Moreover, we include
these effects in the  calculation of the heavy quark production in
$AA$ collisions, verifying that they cannot be disregarded both in
the estimates of quarkonium suppression and in the initial
conditions of the quark-gluon plasma.

\vspace{1.5cm}



\end{titlepage}

The relativistic collider facilities RHIC and LHC will for the
first time provide the opportunity to systematically study the
physics of hot and ultradense matter in hadron-nucleus ($pA$) and
nucleus-nucleus ($AB$) collisions at energies that are orders of
magnitude larger than  the energies of the current accelerators.
The systematic study of $pA$ collisions at the same energies is
essential to gain insight into the structure of the dense medium
effects. Such effects, as the energy loss and shadowing, are
absent or small in $pp$ collisions, but become increasingly
prominent in $pA$ collisions, and are of major importance in $AA$
reactions. By comparing $pA$ and $AA$ reactions involving very
heavy nuclei, one may be able to distinguish basic hadronic
effects that dominate the dynamics in $pA$ collisions, from a
quark-gluon formation predicted to occur in heavy ion $AA$
collisions. To gain insight into the underlying hadronic
processes, one has to study collisions that are expected to not
lead to a QGP formation. Once the physics of ''QCD at high
densities'' is better understood, the mechanisms of quark-gluon
plasma formation and related collective phenomena in heavy ion
collisions could be disentangled from the basic hadronic effects.

One of the nuclear medium effects is the nuclear shadowing, which
is the modification of  the target parton distributions so that
$xq^A(x,Q^2) \, < \,  Axq^N(x,Q^2)$, as expected from a
superposition of $pp$ interactions. The current experimental data
presenting nuclear shadowing can be described reasonably using the
DGLAP evolution equations \cite{dglap} with adjusted initial
parton distributions \cite{eks}. However, this parameterization
does not include the dynamical saturation effects of the parton
distributions predicted at small $x$ and large $A$ \cite{muesat}.
In this kinematical regime, the density of quarks and gluons
becomes very high and the processes of interaction and
recombination between partons, not present in the DGLAP evolution,
should be considered.
 Recently, we have
proposed a procedure to improve the nuclear parton distributions
and include the perturbative high density effects in the
predictions of the inclusive observables in $eA$ processes
\cite{ayavic}.  There we have indicate that the high density
effects becomes very important mainly for the LHC kinematic region
and that these effects  could not  be disregarded in the
calculations of the observables and signatures of the QGP. Here we
analyze in detail the heavy quark production in $pA$ processes in
order to investigate the presence and magnitude of the high
density effects (For a  related discussion of heavy quark
production see Refs. \cite{vogtliuti,greiner}). Our analyzes is
motivated by the fact that the heavy quark production at RHIC and
LHC energies is dominated by initial state gluons. Thus, the $pA$
process becomes a valuable search of information about the gluon
distribution in the nucleus. Since the probability for making a
heavy quark pair is proportional to square of gluon distribution,
any depletion in number of gluons will make a significant
difference in the number of heavy quark pairs produced.

Lets start from a brief review of the high density scenario. At
high densities we expect  the limitation on the maximum
phase-space parton density that can be reached in the hadron
wavefunction (parton saturation) and very high values of the QCD
field strength squared $F^2_{\mu \nu} \propto 1/\alpha_s$
\cite{muesat}, possibly forming a Color Glass Condensate
\cite{iancu}, which is characterized by a bulk momentum scale
$Q_s$. If this saturation scale is larger than the QCD scale
$\Lambda _{QCD}$, then this system can be studied using weak
coupling methods. At present, there are different approaches to
describe hdQCD effects, all of them including nonlinear terms in
parton evolution equations \cite{glr}. In fact, these approaches
sum powers of the probability of gluon-gluon interaction inside of
the parton cascade.   As a summary, we have that  the different
approaches match the DLA limit of the DGLAP parton evolution in
the low parton density limit and match the Glauber-Mueller
approach in the transition regime of low density to the high
density limit. At the energies of interest, we believe that
 the predictions of the Glauber-Mueller approach
can be taken as good approximation to the description of high
density effects. As this approach is a common limit  to the
distinct approaches, we intend to obtain predictions which are not
model dependent.

The main properties of the nuclear gluon distribution  $xG^A (x,
Q^2)$ in the Glauber-Mueller approach  were extensively discussed
in  Ref. \cite{ayala1}. For  completeness, we now present a
qualitative discussion of the main properties of this approach. In
the nucleus rest frame we can consider the interaction between a
virtual colorless hard probe and the nucleus via the gluon pair
($gg$) component of the virtual probe. The interaction of the
dipole with the color field of the nucleus will clearly depend on
the size of the dipole. If the separation of the $gg$ pair is very
small (smaller than the mean separation of the partons), the color
field of the dipole will be effectively screened and the nucleus
will be essentially 'transparent' to the dipole. At large dipole
sizes, the color field of the dipole is large and it interacts
strongly with the target  and is sensitive both to its structure
and size. More generally, when the parton density is such that the
nucleus becomes 'black' and the interaction probability is unity,
the dipole cross section saturates and  the gluon distribution
becomes proportional to the virtuality of the probe $Q^2$. In the
infinite momentum frame, this picture is equivalent to a situation
in which the individual partons become so close that they have a
significant probability of interacting with each other before
interaction with the probe. Such interactions lead, for instance,
to two $\rightarrow$ one branchings and hence a reduction in the
gluon distribution. These properties were considered in Ref.
\cite{ayala1}, where the rescatterings of the gluon pair inside
the nucleus were estimated using the Glauber-Mueller approach,
resulting  that the nuclear gluon distribution is given by
\cite{ayala1}
\begin{eqnarray}
xG_A(x,Q^2) = \frac{2R_A^2}{\pi^2}\int_x^1
\frac{dx^{\prime}}{x^{\prime}}
\int_{\frac{1}{Q^2}}^{\frac{1}{Q_0^2}} \frac{d^2r_t}{\pi r_t^4}
\{C + ln(\kappa_G(x^{\prime}, r_t^2)) + E_1(\kappa_G(x^{\prime},
r_t^2))\} \label{master}
\end{eqnarray}
where $C$ is the Euler constant,  $E_1$ is the exponential
function, the function  $\kappa_G(x, r_t^2) = (3
\alpha_s\,A/2R_A^2)\,\pi\,r_t^2\,
 xG_N(x,\frac{1}{r_t^2})$,
 $A$ is the number of nucleons in a nucleus and $R_A^2$ is the mean nuclear radius.
One of the shortcomings of this approach is that the
Glauber-Mueller formula disregards any nuclear effect in the
nonperturbative initial condition for the gluon distribution.
Furthermore, the medium effects present at larger values  of $x$,
the antishadowing and the EMC effect, are also disregarded in this
approach. In order to improve the Glauber-Mueller approach, it was
recently proposed in Ref.\cite{ayavic} a modification in the
expression (\ref{master}) that includes the full DGLAP kernel in
parton evolution. Basically, we propose to calculate the nuclear
gluon distribution using the following procedure:
\begin{eqnarray}
xG_A(x,Q^2) & = & (1/A)xG_A(x,Q^2)[GM] - (1/A)xG_A(x,Q^2)[DLA]
\nonumber\\
&+& \, xG_A(x,Q^2)[EKS] \,\,\,, \label{proc}
\end{eqnarray}
where $xG_A(x,Q^2)[GM]$ represents the Glauber-Mueller nuclear
gluon distribution given by the expression (\ref{master}) and
$xG_A(x,Q^2)[DLA]$ is the DGLAP (DLA) prediction for the nuclear
gluon distribution, which correspond to the first term of
expression (\ref{master}) when expanded in powers of $\kappa_G$.
The last term in expression (\ref{proc}), $xG_A(x,Q^2)[EKS]$, is
the gluon distribution solution of the DGLAP equation as proposed
by Eskola, Kolhinen and Salgado(EKS) \cite{eks}, where the initial
conditions of the parton evolution are chosen in such a way to
describe the nuclear effects in DIS and Drell-Yan fixed nuclear
target data. As it was discussed in detail in Ref.\cite{ayavic},
the parameterization of the nuclear effects in the gluon
distribution, given by the ratio $R_G^A(x,Q^2) =
xG_A(x,Q^2)/xG_N(x, Q^2)$ in the EKS procedure, does not include
the perturbative high density effects  at small values of $x$.
Thus, the procedure presented in equation (\ref{proc}) includes
the full DGLAP evolution equation in all kinematic region, take
into account the nuclear affect in the present fixed target data
and includes the high density effects in the parton evolution at
small $x$ in the perturbative regime.

To illustrate our results, in Fig. \ref{fig1} we compare our
predictions for the $x$ - dependence of the  ratio $R_G =
xG_A/xG_N$ for $Q^2 = 15$ GeV$^2$ with the EKS predictions. The
results are shown for two typical nuclei of interest in nuclear
collisions. Also the corresponding kinematic region which will be
explored in the RHIC and LHC colliders are presented. Two
important features of the high density effects can be seen in Fig.
\ref{fig1}: the magnitude of the effects and the saturation
behavior. The suppression of the gluon distribution in respect to
the EKS set is 6\% for A=40 (Ca) and 11\% for A=208 (Pb) in the
lower limit of the RHIC kinematic range. When the lower limit of
the LHC kinematic range is concerned, the reduction of the gluon
distribution is 31\% for A=40 and 50\% for A=208. This strong
effect also modify the saturation of the ratio predicted in the
EKS parameterization at low $x$. Basically, in the EKS approach
the saturation of the ratio is assumed in the nonperturbative
initial condition as a constraint for the behavior of the nuclear
parton distributions in the small $x$ limit. This behavior is
preserved by the DGLAP evolution, since these equations reduce to
the DLA limit at low $x$ in the nucleon and nuclear case, keeping
the ratio constant. When the high density effects are considered
in the dynamics,  a large modification of the gluon distribution
is predicted, which is amplified in nuclear processes since the
nuclear medium amplifies the effects associated to the high parton
density. Therefore, the high density approaches predict larger
effects in the nuclear gluon distribution than the nucleon one,
independent of the initial parton distributions used. This
expectation is present in the behavior of $R_G^{A}(AG)$, which
turn to be much smaller then $R_G^{A}(EKS)$ in low $x$ region. As
the heavy quark production is dominated by the fusion of gluons,
we expect that these effects will strongly modify the
$c\overline{c}$ and $b\overline{b}$ production in $pA$ processes
in the high energy limit.

Now let us consider the heavy quark  production. Perturbative QCD
calculations of heavy quark production at leading order have long
been available. For high energies and at leading order (LO),  the
production is dominated by gluon fusion, with charm and bottom
quarks  produced basically in the process $gg \rightarrow
Q\overline{Q}$ with $Q = c$ and $b$. The LO cross section for a
proton $p$ colliding with a nucleus $A$ is then:
\begin{eqnarray}
\sigma_{pA} =\int d\tau \,\int d x_F\, & & \frac{1}{s}\,
\frac{1}{\sqrt{x_F^2 + 4\tau}} \,xG_p(x_1,\mu^2)
\,xG_A(x_2,\mu^2)\, \times \nonumber \\
& &\hat{\sigma}_{gg \rightarrow Q
\overline{Q}}(x_1,x_2,m_Q,\mu^2)\,\,,\label{cross}
\end{eqnarray}
where $x_F=x_1 - x_2$, $\tau=M^2/s$ and $M^2$ is the invariant
mass of the virtual gluon in the subprocess.  $xG_i$ the gluon
distributions evaluated at momentum fraction $x$ and  momentum
scale $\mu^2$. For practical purpose, we take $\mu^2=4 m_c^2$ for
charm production and $\mu^2=m_b^2$ for bottom production. In order
to investigate the medium dependence of the heavy quark production
cross section, we will follow the usual procedure  used  to
describe the experimental data on nuclear effects in the hadronic
quarkonium production \cite{alfass}, where the atomic mass number
$A$ dependence is parameterized by $\sigma_{pA} = \sigma_{pN}
\times A^{\alpha}$. Here $\sigma_{pA}$ and $\sigma_{pN}$ are the
particle production cross sections in proton-nucleus and
proton-nucleon interactions, respectively. If the particle
production is not modified by the presence of nuclear matter, then
$\alpha = 1$. A number of experiments have measured a less than
linear $A$ dependence for various processes of production, which
indicates that the medium effects cannot be disregarded. As our
focus is to analyze the influence of the high density effects in
the heavy quark production when compared to the usual DGLAP-EKS
predictions, we do not consider the contributions of  energy loss
and the intrinsic heavy-quark components for the non-linear $A$
dependence of the cross sections (For a discussion of these
effects in charmonium production see Ref. \cite{vogtxf}). In this
paper we will evaluate the heavy quark production at leading order
and will use the quark masses $m_c = 1.2$ GeV and $m_b=4.75$ GeV.
To estimate the modification of heavy quark production cross
section due to the high density effects, we calculate the
effective exponent $\alpha$, which is given by
\begin{eqnarray}
\alpha =  \left( \left. ln \frac{\sigma_{pA}}{ \sigma_{pN}}\right/
{ln A}\right)\,\,. \label{alfa}
\end{eqnarray}
We calculate the above expression considering, as  an input in the
nuclear cross sections, the  parameterization of the nuclear
effects  in the gluon distributions, which is given by $
xG_A(x_i,\mu^2)= R_G(x_i,\mu^2)\,xG_{N}(x_i,\mu^2) $, with $R_G$
presented in Fig. \ref{fig1} and  $xG_N$  taken from the GRV94(LO)
set \cite{grv95}.  Two  comments are in order here. First, since
we are interested in the behavior of the effective exponent, we
may expect that our results will not be modified by the NLO
corrections (the $K$-factor) \cite{mangano}. Second, we are
assuming that the collinear factorization is still valid in the
kinematic region considered. This is a strong assumption which
should be tested by the comparison of our results with a
calculation of this process considering the color glass condensate
formalism, similarly to made in Ref. \cite{gelis} for
photoproduction. The agreement between the results will allow us
to use our approach as  a simplified method of inclusion of the
high density effects in the nuclear cross sections. Furthermore,
the comparison between the results  of our approach and the color
glass predictions for the gluon minijet production \cite{dumitru}
will be other important cross check of our approach. Work in this
direction is in progress.

 In order to gain insight into the
amount of the high density effects in heavy quark production at
future colliders, we will consider the process calculated at the
proposed energies for RHIC and LHC. These energies are
$\sqrt{s}=200$ GeV per nucleon pair for Au+Au collisions and
$\sqrt{s}=350$ GeV for $pA$ process at RHIC. For LHC,
$\sqrt{s}=5500$ GeV per nucleon pair for Pb+Pb collisions and
$\sqrt{s}=8800$ GeV for $pA$ processes \cite{cmsno}. In Fig.
\ref{fig2} we present the effective exponent $\alpha$ as a
function of the c.m. energy $s^{1/2}$ for two different nucleus.
We verify that the high density effects are sizeable at high
values of energy  even for  a small value of the mass number
($A=40$). The general behavior of the exponent can be understood
as follows. When the integrations in Eq. (\ref{cross}) are taken,
the nuclear gluon distribution are evaluated in the $x_2$ interval
given by $\tau \, < \, x_2 \, < \, \sqrt{\tau} $. Thus, when the
energy grows, the $x_2$ interval goes to the small $x$ region. For
charm production, for example, the antishadowing region in Fig.
\ref{fig1} dominates the integration for energies smaller than 80
GeV. For bigger energies, the shadowing region dominates and the
exponent is smaller than 1. In  bottom production, the $x_2$
interval is dislocated to larger values of $x$, which implies that
the antishadowing region of Fig. \ref{fig1} dominates the
production at energies in the interval 100-200 GeV and $\alpha$ is
close to one even for RHIC energies. For bigger values of energy,
the suppression is sizeable and the effective exponent is smaller
than 1. In general grounds we have that the saturation of the
exponent observed in the EKS prediction is associated to the
behavior of the ratio $R_G$ at small $x$, while the presence of
the high density effects (nonsaturation of $R_G$) implies a large
reduction of the $pA$ cross section when compared with DGLAP-EKS
description of nuclear effects. Therefore, we believe that the
analyzes of the effective exponent in $pA$ processes can be useful
to evidence the high density effects.

In Fig. \ref{fig3}, we present the exponent $\alpha$ for RHIC and
LHC energies as a function of $x_F \equiv x_1-x_2$.  The exponent
was obtained from the expression
\begin{eqnarray}
\alpha(x_F) =  \left\{ \left. ln \left(\left.{\frac{d
\sigma_{pA}}{d x_F}}\right/ {\frac{ d \sigma_{pN}}{d x_F}}
\right)\right/ {ln A}\right\}\,\,, \label{alfaxf}
\end{eqnarray}
where the differential cross section was obtaining integrating
equation (\ref{cross}) in the invariant mass of the virtual gluons
from the heavy quark threshold to the energy squared $s$, which
corresponds to $\frac{m_Q^2}{S} < \tau < 1$.  For values of $x_F$
close to zero, $x_2 \approx x_1$ and the $\tau$ integration
corresponds to a large interval in both $x_2$ and $x_1$. For $x_F$
close to one, both $x_2$ and $x_1$ interval are much smaller, with
$x_1$ close to one and $x_2$ close to zero. The lower limit of
$\tau$ corresponds also to the lower $x_2$ limit. Considering the
kinematics discussed above and the properties of the Fig.
\ref{fig1}, all the features of Fig. \ref{fig3} can be understood.
For the $p-Pb$ processes at RHIC ($\sqrt{s} = 350$ GeV) with $x_F$
close to zero, the exponent is close to one. This occurs because
the $\tau$ integration corresponds to a large interval in $x_2$,
including contribution  from the antishadowing region of Fig.
\ref{fig1}. For  charm production, the threshold is low enough to
include the contribution from the small $x_2$ region and the
exponent $\alpha$ is smaller than one. For the bottom production,
the low $x_2$ contribution is small and the exponent is bigger
than one. Since high density effects are small in the intermediate
$x_2$ region, the EKS and AG prediction are close to each other.
For $x_F$ close to one, the low $x_2$ behavior of the gluon
distribution dominates the process. Since the EKS parameterization
predicts that the ratio saturates at this region, the exponent
tend to a constant, while the high density effects predict an
exponent that decrease monotonically. These effects are much
stronger in the charm production due to the lower threshold.
Moreover, all effects discussed above are amplified for LHC energy
($\sqrt{s}= 8.8$ TeV), even for $x_F$ close to zero. Due to the
large energy, the lower limit of the $\tau$ integration is very
close to zero. Then, the low $x_2$ contribution is much bigger and
the high density effects are much more important. For $x_F$ close
to one, the EKS prediction indicates a constant value for
$\alpha$, both for charm and bottom production. This is a direct
consequence of the saturation of the ratio $R_G$ in the low $x_2$
region. As far as the high density effects are considered, the
exponent is strongly reduced. As $x_F$ tend to one, $\alpha$ goes
to $0.6$ for charm production and $0.7$ for bottom production.

The strong modification of the heavy quark production in $pA$
process indicate that high density effects on the gluon
distribution will play an important role in the calculation of the
initial conditions and signatures of the quark-gluon plasma in
heavy ion collisions. To illustrate this point, we present in Fig.
\ref{fig4} the exponent $\beta$ calculated by the expression
\begin{eqnarray}
\beta = \left( \left. {ln \frac{\sigma_{AA}}{
\sigma_{pN}}}\right/{ln A}\right)\,\,, \label{beta}
\end{eqnarray}
for the heavy quark production in Pb-Pb collisions as a function
of $\sqrt{s}$. The exponent present a similar feature to the
exponent $\alpha$ shown in Fig. \ref{fig2}. For energies close and
smaller than $100$ GeV, the most important contribution comes from
the antishadowing region of Fig. \ref{fig1}, and the exponent is
bigger than two. For larger energies, the small $x$ region gives
the more important contribution. When compared to the EKS
prediction, the high density effects strongly reduces the
production cross sections. At RHIC energy ($\sqrt{s}= 200 GeV$),
the exponent $\beta$ is reduced in 8\% for $c \bar c$ and 4\% for
$b \bar b$. At LHC energy ($\sqrt{s}=5500$), the relative
reduction is 42\% for $c \bar c$ and 28\% for $b \bar b$.  As we
can see, the high density effects will strongly  reduce the
production cross sections in $AA$ processes and should be taken
into account to calculate the cross section of the hard QCD
process in the first stage of the heavy ion collisions.

In brief, a systematic measurement of gluon shadowing is of
fundamental interest in understanding of the parton structure of
nuclei as well as  in the field of minijet production that
determines the total entropy produced at RHIC and higher energies.
In this paper we addressed the heavy quark production in $pA$
processes as a search to verify the presence and estimate the
magnitude of the high density effects and, consequently, fix the
behavior of the gluon distribution at large energies. In this
sense our results can be considered as complementary to the
predictions made by
 Wang and Gyulassy
 \cite{gyu}, which have studied the sensitivity of
single-particle inclusive spectra in nuclear collisions to gluon
shadowing and jet quenching, and have suggested the investigation
of  the $pA$ collisions at the same energy in order to disentangle
both effects. Moreover, the study of heavy quark production is
important to estimate   the suppression of this process associated
to the high density effects, which is necessary for a reliable
calculation of quarkonium production and its suppression in a
quark-gluon plasma. As a summary of our results we have calculated
the cross sections for heavy quark production in $pA$ processes
and estimated the energy and $x_F$ dependencies of the effective
exponents that parameterize the medium effects. Our results
demonstrate that these effects are large even at small nuclei and
that a systematic experimental analyzes  could discriminate
between the high density regime and the predictions from the
linear regime. Furthermore, we have extended our analyzes for
heavy quark production in heavy ion collisions and estimated the
contribution of the high density effects in this process. Our
results have important implications in the signatures of the QCD
phase transition, specially in the suppression of quarkonium
production since our results  indicate that the quarkonium
production rate should be strongly modified by the presence of the
high density effects. Consequently, if the analyzes of the $pA$
process demonstrate the presence of the these effects  in the
kinematic regions of RHIC and LHC, we have that the current
estimates of quarkonium suppression in $AA$ processes should be
completely reanalyzed.

\section*{Acknowledgments}
The authors acknowledge helpful discussions with M. B. Gay Ducati,
J. Jalilian-Marian, M. V. T. Machado and R. Venugopalan. VPG
thanks the Brookhaven National Laboratory for its hospitality at a
preliminary stage of this work. This work was partially financed
by CNPq and FAPERGS, BRAZIL.

\newpage

\section*{Figure Captions}

\begin{figure}[tbp]
\centerline{\psfig{file=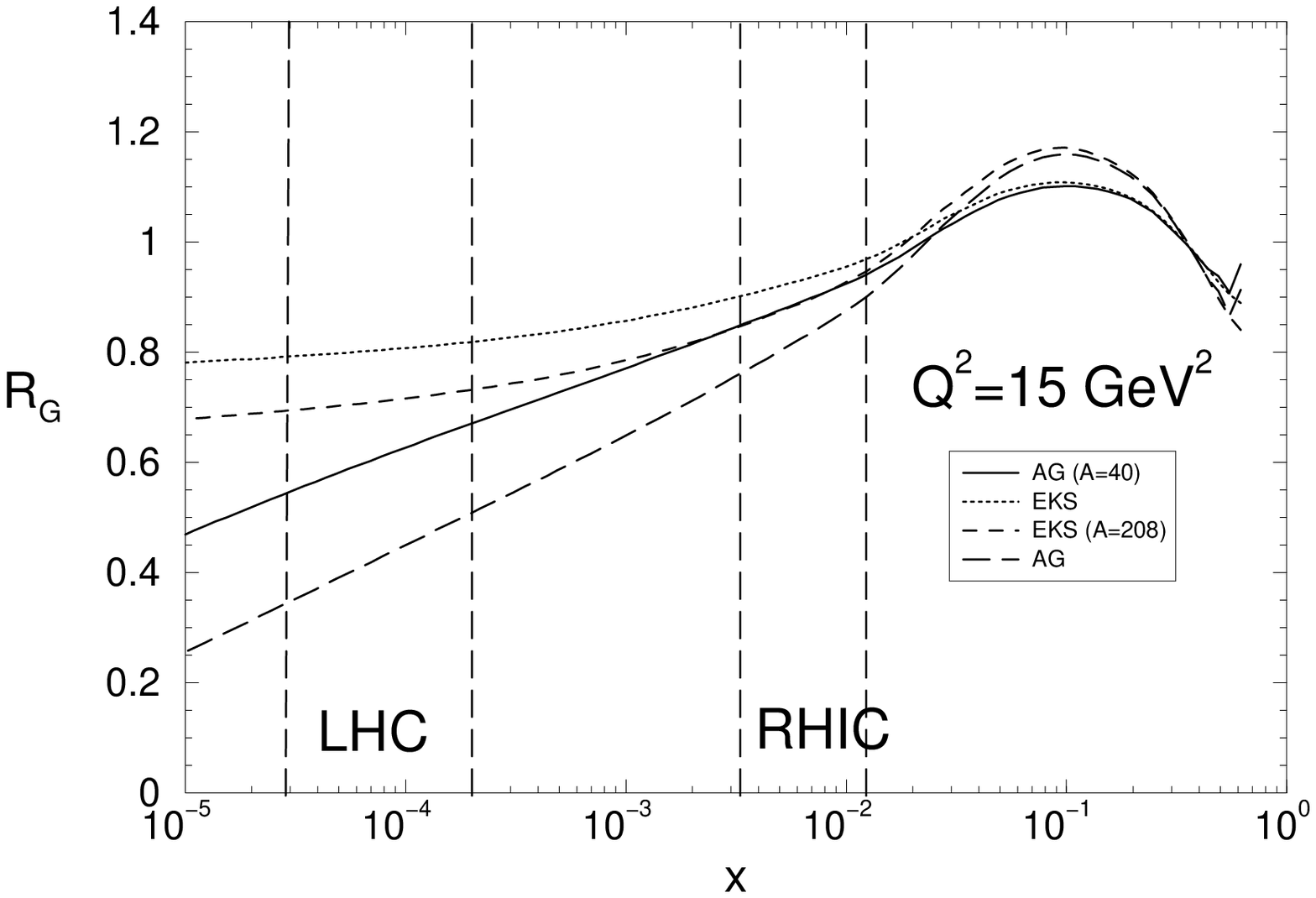,width=150mm}} \caption{}
\label{fig1}
\end{figure}

\vspace{1.0cm} Fig. \ref{fig1}: Comparison between the ratio $R_G$
with and without the high density effects for A=40 (Ca) and A=208
(Pb). The RHIC and LHC kinematic regions are also shown.

\begin{figure}[t]
\begin{tabular}{c}
\psfig{file=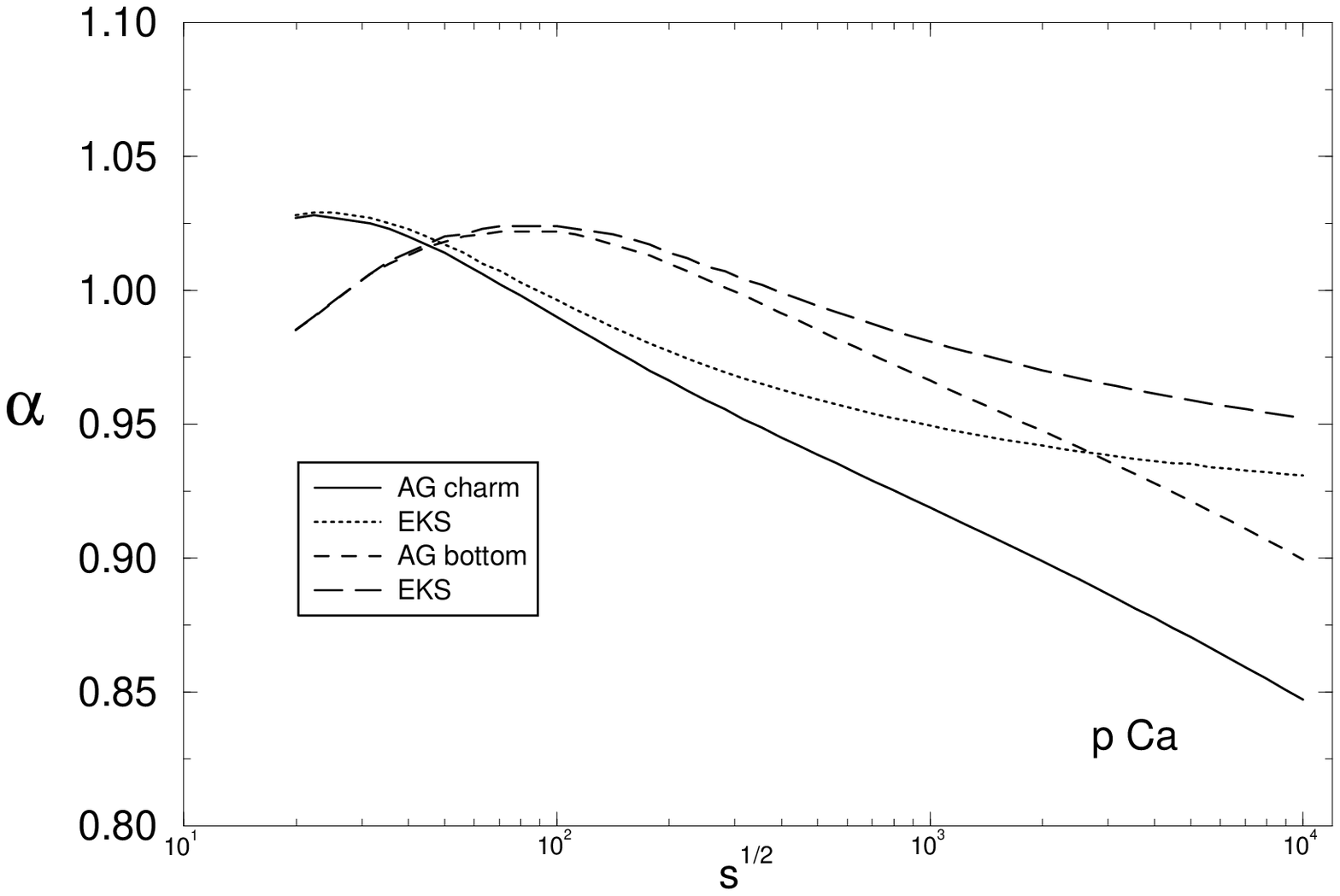,width=120mm} \\
\psfig{file=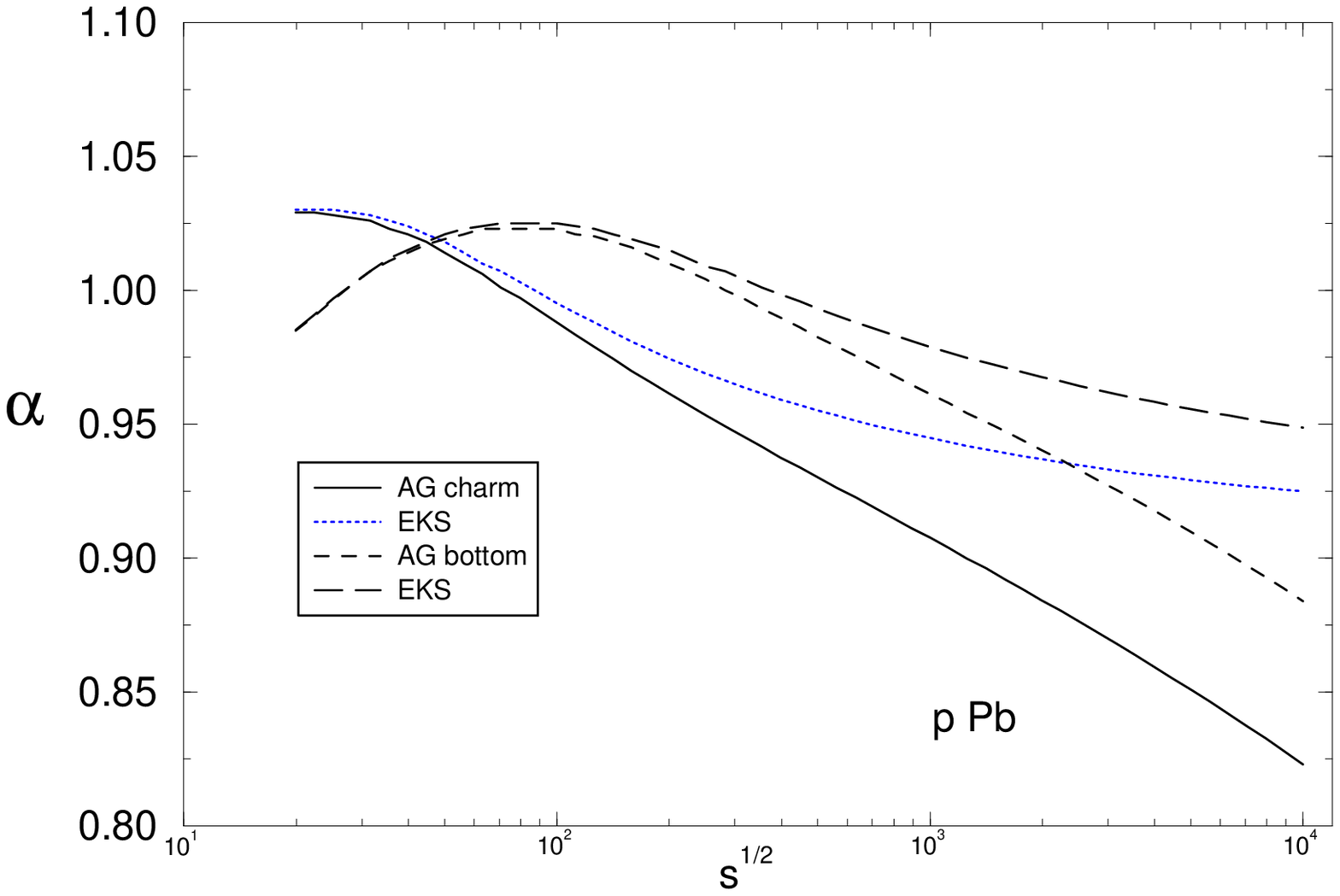,width=120mm}
\end{tabular}
\caption{} \label{fig2}
\end{figure}

\vspace{1.0cm} Fig. \ref{fig2}: The exponent $\alpha$ [Eq.
(\ref{alfa})] as a function of energy $\sqrt{s}$ for two nuclei
(Ca and Pb) in $pA$ processes.

\begin{figure}[t]
\begin{tabular}{c}
\psfig{file=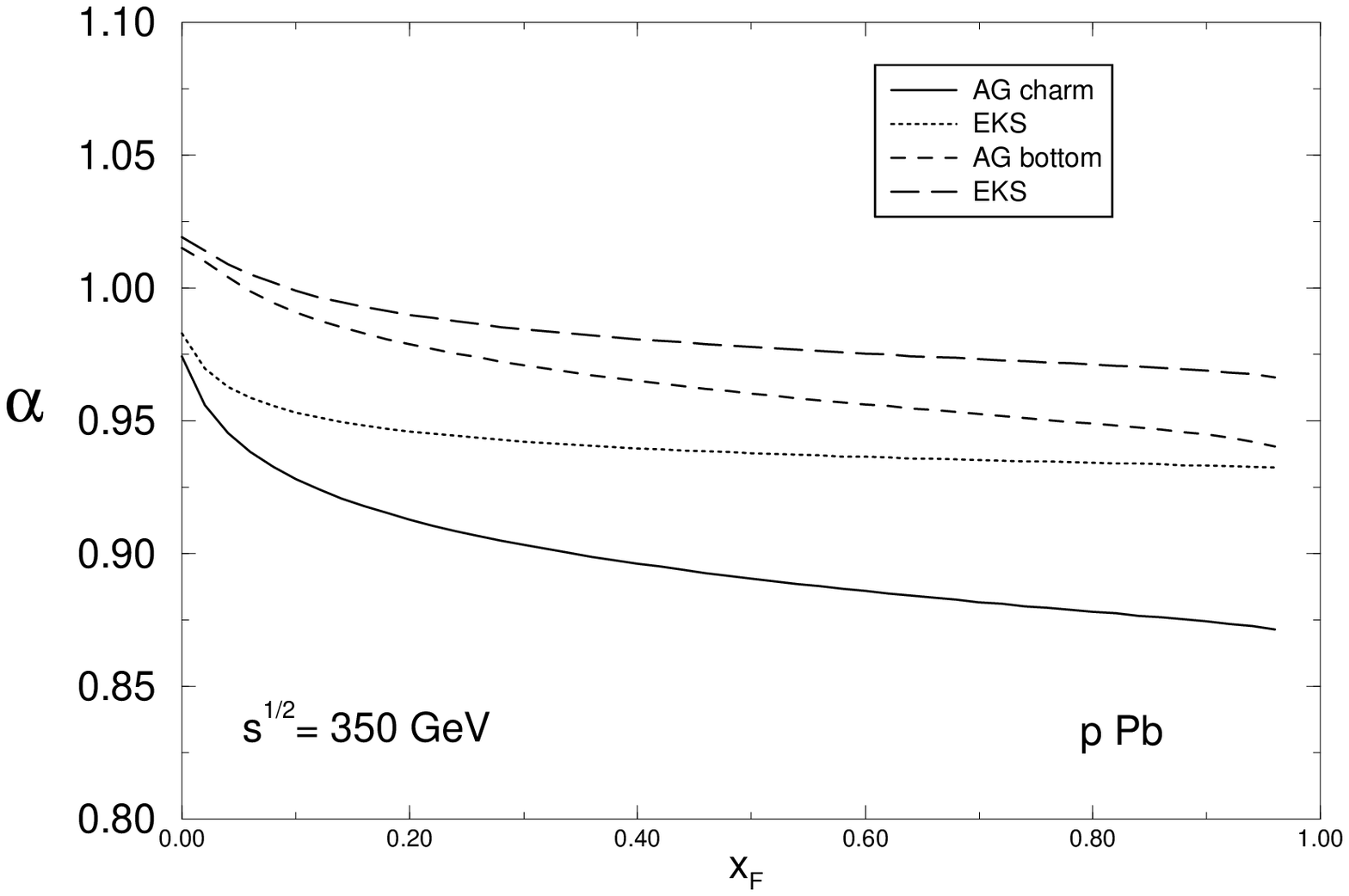,width=120mm} \\
\psfig{file=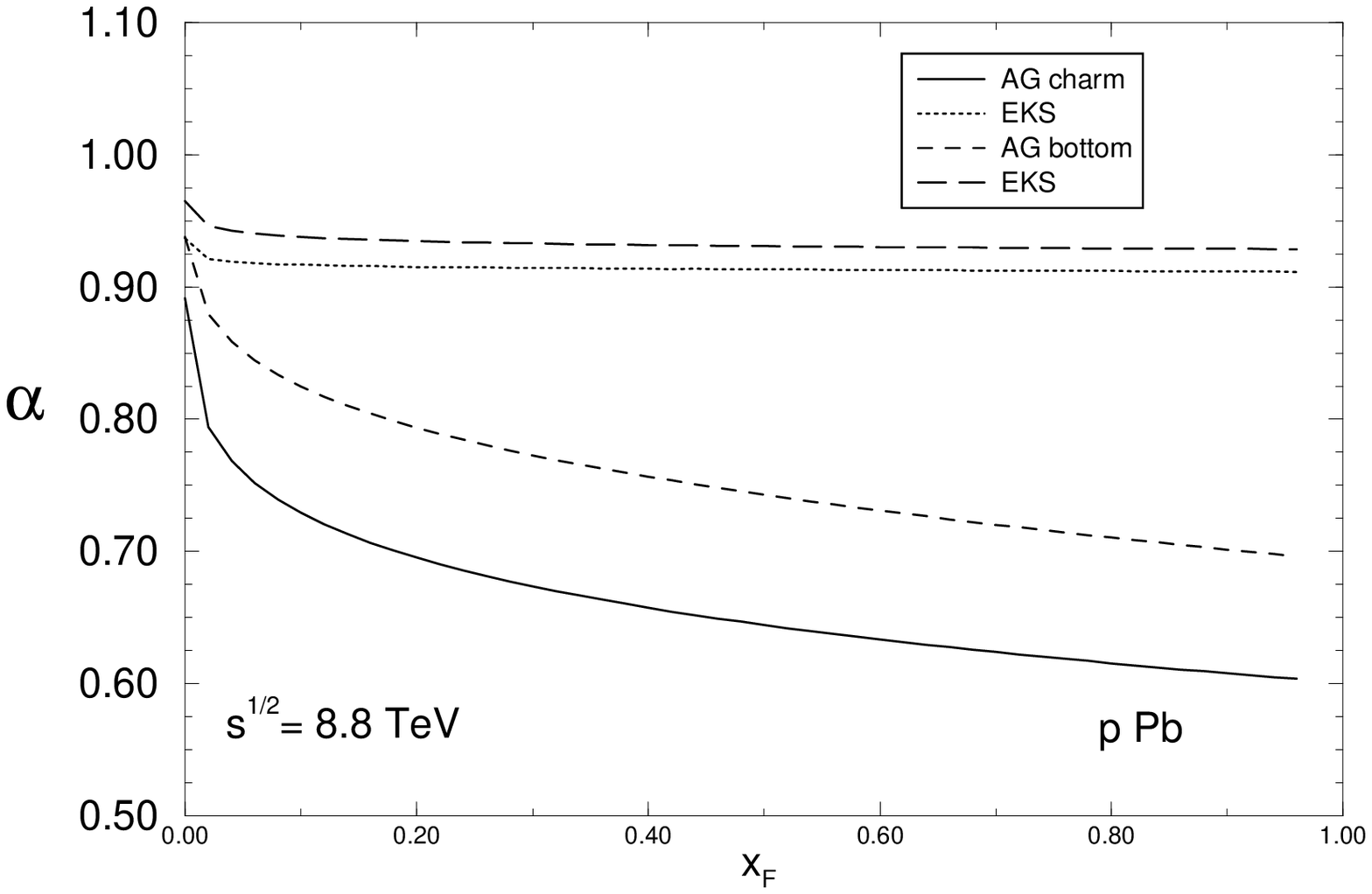,width=120mm}
\end{tabular}
\caption{} \label{fig3}
\end{figure}

\vspace{1.0cm} Fig. \ref{fig3}: The exponent $\alpha(x_F)$ [Eq.
(\ref{alfaxf})] as a function of $x_F$ for $p-Pb$ processes at
RHIC ($\sqrt{s}= 350$ GeV) and LHC ($\sqrt{s}=8.8 $ TeV) (A=208)
[Note  the different inferior limit of the two graphs].

\begin{figure}[tbp]
\centerline{\psfig{file=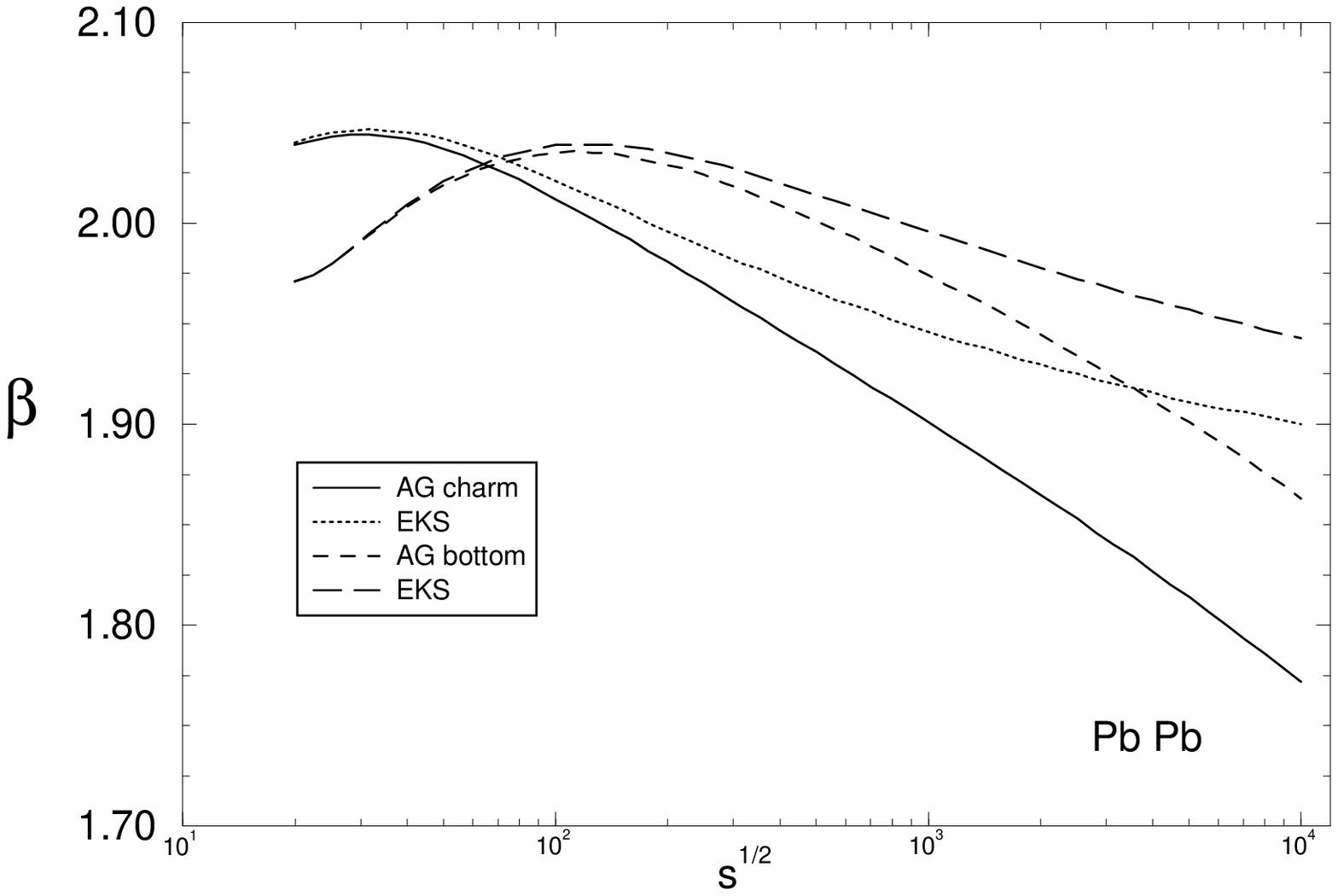,width=150mm}} \caption{ }
\label{fig4}
\end{figure}

\vspace{1.0cm} Fig. \ref{fig4}: The exponent $\beta$ [Eq.
(\ref{beta})] as a function of energy $\sqrt{s}$ for $Pb-Pb$
processes.

\newpage





\end{document}